\documentclass{ragtime}

\usepackage{float}
\usepackage{amsthm}
\usepackage[mathscr]{euscript}
\usepackage{times}
\usepackage{txfonts}
\usepackage{color}
\usepackage{dirtytalk}
\usepackage[T1]{fontenc}
\usepackage[utf8]{inputenc}
\usepackage[english]{babel}

\usepackage{soul,ulem}

\definecolor{darkgreen}{rgb}{0.0, 0.75, 0.0}

\title[Numerical Integration of Geodesic Motion in Non-Schwarzschild Spacetimes]
      {Numerical Integration of Geodesic Motion in Non-Schwarzschild Spacetimes}

\author[Jan Pijáček, Georgios Lukes-Gerakopoulos, Ondřej Zelenka
            ]
       {Jan Pijáček\ins{1,a}, Georgios Lukes-Gerakopoulos \ins{2,}\ins{~3}, Ondřej Zelenka\ins{2} \\ 
        \\
        \ins{1} \hspace{-0.5mm}Institute of Theoretical Physics,
       Faculty of Mathematics and Physics, Charles University,\splitins[1]
        \hspace{-0.5mm} V Holešovičkách 747/2, CZ-180 00 Prague,
        Czech Republic\\
        \ins{2} \hspace{-0.5mm}Astronomical Institute,
       Czech Academy of Sciences,\splitins[2]
       \hspace{-0.5mm}Boční II 1401/1a, CZ-141 00 Prague,        Czech Republic\\
        \ins{3} \hspace{-0.5mm}Department of Physics, University of Thessaly,
       GR-35100 Lamia, Greece\\
        \ins{a}\hspace{-0.5mm}\Email{89602231@cuni.cz}} 

\begin{document}

	\begin{abstract}

        This contribution presents a method for numerically solving geodesic motion in Wolfram Mathematica applied on a system consisting of a Schwarzschild black hole surrounded by a thin Vogt-Letelier disk. While geodesic motion in the Schwarzschild spacetime is fully integrable due to its high degree of symmetry, the addition of a disk breaks the spherical symmetry and leads to a weak-integrable dynamical system. Particular attention is paid to the treatment of particle trajectories near the thin disk, where discontinuities in the derivatives of the metric may affect numerical stability and the conservation of integrals of motion. Several numerical treatments of disk crossings were implemented, arriving to a method based on the re-normalization of the four-velocity. The phase-space of the system is investigated using Poincar\'{e} sections and rotation number spectra.

	\end{abstract}

	
	\section{Introduction}\label{sec:intro}
Extreme mass ratio inspirals (EMRIs) are binary systems which involve a stellar black hole or a neutron star orbiting a supermassive black hole. The lighter compact object loses energy and angular momentum due to radiation reaction which drives it in an inspiral motion towards the central supermassive black hole. When modeling EMRI systems, in a zero order approximation, the influence of the lighter object is negligible and its motion can be described as a gradual shift through geodesics towards the central object as a result of the gravitational wave emission \citep{Pound2021}. Supermassive black holes are located at the center of galaxies and they are accreting matter. This motivates the investigation of geodesic motion of a test particle around an astrophysical model that involves a black hole surrounded by matter such as in an active galactic nuclei. Hence, this work studies geodesic motion in an example of such model. Its system consist of a Schwarzschild black hole surrounded by a thin Vogt-Letelier disk
located in the equatorial plane~\citep{Kotlařík_2022}. The corresponding metric is a vacuum solution
of the Einstein field equations. The disk represents a perturbation of otherwise
integrable Schwarzschild metric leading to a chaotic behaviour of the test particle.
Particular attention is paid to the trajectory crossing of the disk.

\section{Notation and conventions}

Throughout this work, adapted (geometrized) units are used in order to simplify expressions by setting certain physical constants (e.g. $G$, $c$) to unity. For geodesic motion, the masses can be absorbed by redefining properly the quantities with their dimensionless counterparts, e.g. the radius as $r/M$, the energy as $E/m$, the angular momentum as $L_z/(m M)$. This allows us, also to set in our numerical computations $M=m=1$, essentially ignoring the masses in our study. For the metric components, it is important to note that the signature convention $(-,+,+,+)$ is used. 

\section{The metric used}

Any static, axially symmetric vacuum spacetime can be described by a Weyl-type metric~\citep{Bicak:GTRtextbook}.  
\begin{align}
ds^2 = -e^{2\nu}\,dt^2 + {\rho}^{2}e^{-2\nu}\,d{\phi}^2 + e^{2\nu-2\lambda}(d{\rho}^2 + d{z}^2) .
\end{align}
The line element is expressed in Weyl-Lewis-Papapetrou coordinates, where ${\rho}$ is a cylindrical-type radius and $z$ is a linear coordinate orthogonal to the equatorial plane. The specific spacetime is given by the metric functions ${\nu}$ and ${\lambda}$. The function ${\nu}$ corresponds to the Newtonian potential and is additive through subsystems, ${\lambda}$ is a metric function expressing curvature in the meridional (${\rho}$ - $z$) plane. The system studied here consists of a Schwarzschild black hole surrounded by a thin Vogt-Letelier disk located in the equatorial plane. The Vogt-Letelier disks are obtained by superposition within the inverted Kuzmin-Toomre disks \citep{Kotlařík_2022}. The disk represents a surface mass density distribution. The metric potentials depend solely on $\rho$ and $z$ coordinates. From the Einstein equations follows:
\begin{align}
&\nu_{,\rho\rho} + \frac{\nu_{,\rho}}{\rho} + \nu_{,zz} = 0,\\
&\lambda_{,\rho} = \rho \left( \nu_{,\rho}^2 - \nu_{,z}^2 \right), \quad
\lambda_{,z} = 2\rho \nu_{,\rho} \nu_{,z}.
\end{align}

Due to the additivity of $\nu$, the corresponding metric function of the system can be expressed as $\nu = \nu_{Schw} + \nu_{disk}$. On the other hand, $\lambda$ has to include an interaction term: $\lambda = \lambda_{Schw} + \lambda_{disk} + \lambda_{int}$, where the interaction term is given by:

\begin{align}
\lambda_{\text{int},\rho} &= 2\rho \left( \nu_{1,\rho} \nu_{2,\rho} - \nu_{1,z} \nu_{2,z} \right), \\
\lambda_{\text{int},z} &= 2\rho \left( \nu_{1,\rho} \nu_{2,z} + \nu_{1,z} \nu_{2,\rho} \right).
\label{lambda_int}
\end{align}

The metric is static and diagonal, reflecting the time symmetry of the system.

\section{Equations of motion}
As mentioned already, in the approximation of negligible mass of the test particle in comparison to the mass of the central object, it is possible to describe the motion by geodesic equations. It means that the particle mass doesn’t affect the metric and its motion can be described as free fall. The particle is still in its local inertial frame (LIFE) which is transformed by parallel transport alongside a trajectory $ x_{\mu} $ \citep{Bicak:GTRtextbook}.
\begin{align}
\frac{d^2 x^{\mu}}{d{\tau}^2 }+{\Gamma}^{\mu}_{\rho\sigma}\frac{dx^{\rho}}{d{\tau}}\frac{dx^{\sigma}}{d{\tau}}=0,
\end{align}
where $\tau$ is the proper time and ${\Gamma}^{\mu}_{\rho\sigma}$ are Christoffel symbols of second type.

\section{Integrals of motion}

In this thesis, the used metric is constructed by metric potentials dependent solely on coordinates $\rho$ and $z$, the Killing vector fields are then along coordinates $t$ and $\phi$. The corresponding conserved quantities are:
\begin{align}
    E &= -g_{tt}{\dot t}\\ 
    L_z &= g_{\phi\phi} {\dot \phi},
\end{align}
where $E$ is the test particle's specific energy and $L_z$ is the specific z-component of it's angular momentum, i.e. energy and angular momentum per mass of the particle.
A metric tensor is trivially also a Killing tensor leading to four-velocity as a conserved quantity
\begin{align}
    g_{\mu \nu} u^{\mu} u^{\nu} = -1.
\end{align}

\section{Integrability in the Hamiltonian formalism} 
In the Hamiltonian formalism, the generalized coordinates and momenta can be chosen to simplify the description of the system. Of particular importance are canonical transformations, which preserve the symplectic structure and the form of Hamilton’s equations. Among these transformations, especially useful is the transformation to action–angle variables~\citep{MeissRevModPhys:1992,Horak2003} $\{\theta^i,J_i\}$, where $\theta^i$ are generalized positions and $J_i$ are generalized momenta defined as:
\begin{align}
J_i \equiv \frac{1}{2\pi} \oint p_idq^i. 
\end{align}
If it is possible to represent the general Hamiltonian $H=H(q^i,p_i,t)$ in the cyclic form $H = H(J_i)$, i.e. dependent only on the momenta, the derivative with respect to $Q^i$ is equal to zero. Therefore, the momenta are constant along the evolution and the expressions for the angles $Q^i$ simply follow from the Hamilton equations:
\begin{align}
\theta^i = \frac{\partial H}{\partial J_i}t +  \theta^i(0) = \omega^i t + \theta^i(0).
\end{align}

The system is integrable and motion is restricted to an N-dimensional torus in the phase space, where the momenta $J_i$ determine the size and ratio of the circles of the torus, whereas angles $\theta^i$ describe the position on the torus. $\omega^i$ are angular velocities of circular motion on the invariant torus. If the Hamiltonian depends only on $N'< N$ of the momenta, the remaining angular velocities are null. The torus then has dimension $N'$ and the motion is equally simple.

\subsection{Non-integrability}
Hamiltonian is called non-integrable if it is not possible to find a transformation that would allow it to be expressed in the cyclic form. Hamiltonian is then a function of both actions $J_i$ and angles $\theta^i$. The special case of such functions is integrable Hamiltonian perturbed by a non-integrable term.

\section{Poincaré section}
Poincaré section is an important instrument for investigation of motion and trajectory behaviour in the phase space as it reduces the dimension of the analyzed space and discretizes the trajectories into points of the section. The Poincaré section is an intersection in the phase space, transversal to the trajectories. The points are typically considered only when passing in a certain direction. In this work, the sections are constructed as intersections of the equatorial plane with trajectories passing from negative to positive values of the  $z$ coordinate. In the analysis of these sections, typical modes of motion can be identified. Periodic trajectories appear as discrete points of the section, if they are stable, they are located at the centers of the islands of stability. Quasi-periodic trajectories, appear as closed curves surrounding the stable periodic points. Both of these types of trajectories belong to resonances and they are called resonant. The area between the islands of stability is called a chaotic sea and is filled with the chaotic trajectories.

\section{Rotation numbers}
Even though a Poincaré section contains adequate information about the motion in a system, for a more thorough analysis dynamical spectra can be applied for localization of the islands of stability and more exact description of the modes of motion of the trajectories on the section.

On a two dimensional map, the angle $\Theta_i$ between the vector pointing to a point of the section from the center $C$, which is usually set to be the local minimum of the effective potential where the center of the main island of stability lies, and a determined axis passing through $C$ is called an azimuth. An angle between two subsequent points on the section
\[
\theta_i = \Theta_{i + 1} - \Theta_i 
\]
is called a rotation angle. The range of the angle is chosen so that no discontinuities arise in the spectrum of the rotation angle \citep{NVoglis_1998,Contopoulos:2002ocda}. In this case $\theta_{i}\,{\in}\,[0,2\pi)$. Rotation number is a limit average value of $N$ subsequent rotation angles \citep{NVoglis_1998,Contopoulos:2002ocda}, i.e.
\[
\nu_{\theta} = \lim_{N\rightarrow\infty}\frac{1}{2\pi N}\sum_{i=1}^{N} \theta_{i}.
\]

When plotted against a parameter (e.g. initial $\rho$ coordinate), the rotation numbers form a rotation curve. In a fully non-degenerate integrable system, the trajectories lie on the surfaces of the invariant tori, each characterized by a unique rotation number. The rotation curve, if crossing every invariant circle only once, is then continuous monotonous curve.

When the system is perturbed, the phase-space structure changes, so only some of the tori remain, deformed under the perturbation, as KAM tori. The KAM curves generate continuous rotation curves and correspond to quasiperiodic orbits. Resonances are filled by quasi-periodic orbits forming islands of stability and chaotic trajectories. For the last mentioned type, the rotation numbers do not evolve smoothly and appear as discontinuities in the rotation curve. The islands of stability are formed by confined points from quasi-periodic trajectories. The points share very similar angular position in the section, so it is possible to identify the island as an interval of effectively constant rotation numbers, which is also disconnected form the rest of the curve.

\section{Numerical solution} \label{sec:NumSol}

To integrate the geodesic equation of motion on the disk the function NDSolve in the program Wolfram Mathematica was employed. The integrator has to be adapted appropriately to calculate the trajectories crossing the disk, which is represented essentially by a delta function on the equator creating a discontinuity there. The suitability of the methods employed will be judged by the conservation of integrals of motion and the correct depiction of the islands of stability on the Poincaré sections. On the sections the trajectories are always set to start with $\dot{ \rho } = 0$.  Due to the reflection symmetry of the spacetime along the equatorial plane, all the trajectories ought to be symmetrical with respect to the $z=0$ axis in the $\rho-z$ plane as well \citep{Witzany_2015}. Hence, the velocities $\dot{\rho}$ of the particle, during crossing of the equatorial plane in a specific $\rho$, have to be of the same absolute value, differing only by the direction, so the Poincaré sections have to be reflection symmetrical with respect to the $\dot{\rho} = 0$ axis.

\subsection{Reflection}

Because of the singular presence of the disk, the integrator has to be adapted, so the trajectory is not integrated through the disk. The simplest modification of the integrator is to make it jump over the disk and reflect the coordinate $z$, whenever the particle reaches a sufficiently small distance from the disk \citep{Witzany_2015}. We shall refer to this procedure from now on as a reflection. With such condition, there is a danger of the particle ending up positioned in $z$ coordinate closer to the disk, than is the distance chosen in the reflection condition, which would lead to a reflection of the particle to $-z$ being close enough to the disk to be reflected back again. To avoid false multiple reflections around the disk, a condition on direction of a velocity component $\dot{z}$, requiring the particle to move towards the disk, must be met.

The velocity condition is applied as:
\[
\text{sign}(\dot{z}) = -\text{sign}(z).
\]

\begin{figure}[h]
    \centering
    \includegraphics[width=0.85\linewidth]{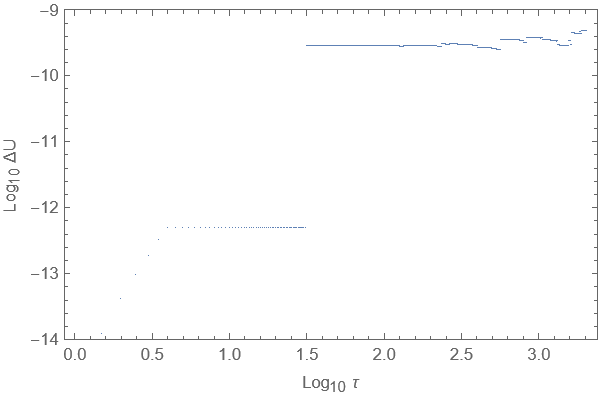}
    \caption{Growth in the relative error of four-velocity for a trajectory generated by the simple reflection.}
    \label{fig:error.growth.reflection}
\end{figure}

Once a crossing is detected, the reflection is simply implemented by
\[
z \rightarrow -z,
\]
while all other coordinates and velocities remain unchanged.  Although simple, this method does not fully control the accumulated numerical error in the integrals of motion as shown in Fig.~\ref{fig:error.growth.reflection}. The position in $\rho$ stays unchanged after reflection, causing a continuation of the trajectory on a slightly different geodesic. Because the metric components depend on $\rho$ and $z$, this disparity between the change in $z$ and no adjustment of $\rho$ leads to an increase of the four-velocity error
\begin{align}
    \Delta U=|1+g_{\mu\nu}u^\mu u^\nu|. 
\end{align}
This shift happens on a scale typically four orders smaller than the range of $\rho$ position in a trajectory, so the importance of the unrealistic treatment is judged based on the growth of the error of the four-velocity.

Figure~\ref{fig:error.growth.reflection} shows the error of four-velocity plotted against the proper time $\tau$. The numerical error, for the integration times used, for a reasonably reliable integration method is not meeting our expectations. It is visible, that the error grows with jumps to the order of $10^{-8}$. This implied that further adjustments of the method are needed for more physically realistic continuation of the trajectory and better conservation of the four-velocity.

\begin{figure}[h]
    \includegraphics[width=0.85\linewidth]{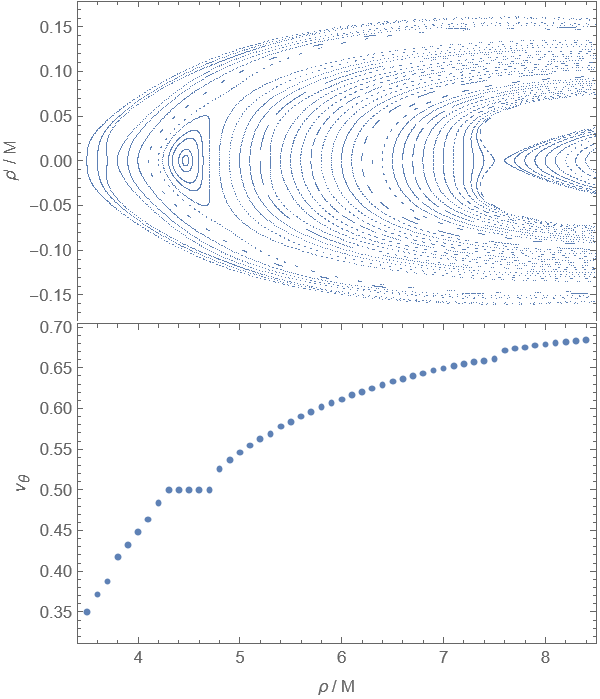}
    \caption{Surface of section for the simple reflection and corresponding rotation curve; parameters taken: $E = 0.95, \, L_{z} = 3.8, \, \mathcal{M}=0.2, \, b=10$ \label{rotnumb1}}
\end{figure}

Figure~\ref{rotnumb1} depicts the detail of the Poincaré section in the range of the initial conditions and the corresponding rotation curve. The first island of stability is clearly visible in the section corresponding to the plateau of value $\frac{1}{2}$ in the rotation curve, disconnected from the rest of the curve. In the center of the island of stability lies a stable point, which is not visible in the section. The chaotic trajectories filling the area between the islands of stability are slightly visible, even though most of the islands are not visible, due to the condition on the direction of crossing through the equatorial plane and the resolution. Approximately in $\rho\,/\,M=3.8$ a jump in the rotation curve around the value of $\nu_{\theta}=\frac{2}{5}$ can be seen. The jump corresponds to the resonance of such value. Similarly, around $\rho\,/\,M=7.6$ a jump in the rotation curve corresponding to the resonance $\frac{2}{3}$ is visible. The jumps appear instead of the plateaus, because the initial conditions correspond to trajectories passing through an unstable point and not an island of stability. For the chaotic layers the rotation numbers would change non-monotonously, even though these parts of the rotation curve are not visible. The structures corresponding to the blank areas in the section indicate islands of stability surrounded by the chaotic layers. The KAM trajectories are visible in the rest of the section with the corresponding continuous part of the rotation curve. On the scale used, the differences between the sections with the corresponding rotation curves for different methods are not visible. For this reason a comparison of a detail of an island of stability is shown later on in the thesis.

\subsection{Linear Prediction}

To improve the integrator, so that the trajectory continues more realistically, reflection can be supplemented with a linear correction of the coordinate $\rho$. When the particle enters the reflection region, the quantities $\dot{\rho}$, $z$, and $\dot{z}$ are recorded. Approximating the trajectory by a locally linear evolution near the plane, the coordinate time passed can be expressed as:

\[
\Delta t = \frac{2z}{\dot{z}}.
\]

The corrected reflection then becomes
\[
z \rightarrow -z, \qquad
\rho \rightarrow \rho + \dot{\rho}\,\frac{2z}{\dot{z}} .
\] 


\subsection{Four-velocity Correction}

Furthermore a reduction of the error growth can be achieved by renormalizing the four–velocity after every 
crossing. Since the integrals of motion $E$ and $L_z$ are known, one can reconstruct 
$\dot{t}$ and $\dot{\phi}$ according to
\[
\dot{t} = -\frac{E}{g_{tt}}, \qquad
\dot{\phi} = \frac{L_z}{g_{\phi\phi}}.
\]
The remaining components, $\dot{\rho}$ and $\dot{z}$, must satisfy the normalization 
condition
\[
g_{\mu\nu} \dot{x}^{\mu}\dot{x}^{\nu} = -1.
\]

 Because the adjustment of $\rho$ is expected to make a greater contribution to the four-velocity error, as $\dot{ \rho}$ is approximated as constant during the reflection, than the reflection of $z$, which is more controlled, the velocity component $\dot{\rho}$ is reset to the value implied by the normalization:

\[
\dot{\rho} =\text{sign}(\dot{\rho})
\sqrt{
\frac{
-1
- g_{tt}\dot{t}^{\,2}
- g_{\phi\phi}\dot{\phi}^{2}
- g_{zz}\dot{z}^{\,2}
}{g_{\rho\rho}}
}.
\]

Inserting the expressions for $\dot{t}$ and $\dot{\phi}$ yields:
\[
\dot{\rho} = \text{sign}(\dot{\rho})
\sqrt{
\frac{
-1
- g_{tt}\left(\frac{E}{g_{tt}}\right)^{2}
- g_{\phi\phi}\left(\frac{L_z}{g_{\phi\phi}}\right)^{2}
- g_{zz}\dot{z}^{\,2}
}{g_{\rho\rho}}
}.
\]

Unfortunately the lack of control over the $\dot{z}$ component leads to non-numerical values after multiple reflection. Even though only two integrals of motion are available, which makes it impossible to reset all velocity components simultaneously, the problem can be overcome by alternating which component is corrected at each crossing. The alternation is controlled by the sign of $z$, which flips at every reflection. Using this mechanism, the corrections are defined as
\[
\dot{\rho} \rightarrow \dot{\rho} + 
\frac{1}{2}\left(\operatorname{sign}(z)+1\right)
\left[
-\dot{\rho}
+ \text{sign}(\dot{\rho})\sqrt{
\frac{
-1
- g_{tt}\left(\frac{E}{g_{tt}}\right)^{2}
- g_{\phi\phi}\left(\frac{L_z}{g_{\phi\phi}}\right)^{2}
- g_{zz}\dot{z}^{\,2}
}{g_{\rho\rho}} }
\right],
\]
\[
\dot{z} \rightarrow \dot{z} - 
\frac{1}{2}\left(\operatorname{sign}(z)-1\right)
\left[
-\dot{z}
+ \text{sign}(\dot{z})\sqrt{
\frac{
-1
- g_{tt}\left(\frac{E}{g_{tt}}\right)^{2}
- g_{\phi\phi}\left(\frac{L_z}{g_{\phi\phi}}\right)^{2}
- g_{\rho\rho}\dot{\rho}^{\,2}
}{g_{zz}} }
\right].
\]

In this way, $\dot{\rho}$ is corrected when passing from positive $z$ to negative and $\dot{ z }$ when passing from negative to positive $z$, ensuring that both components are regularly controlled. This alternating normalization keeps the four–velocity normalization and both integrals of motion conserved to an accuracy of order $10^{-15}$.

\begin{figure}[H]
    \centering
    \includegraphics[width=0.85\linewidth]{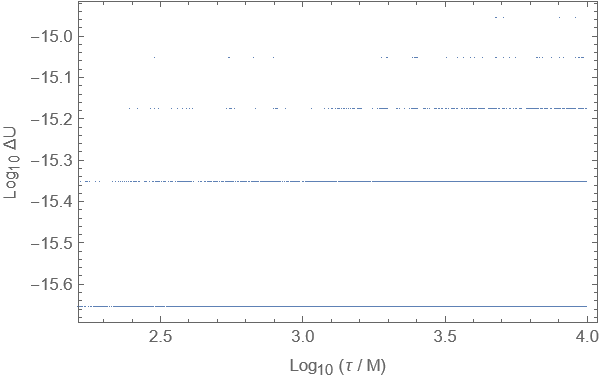}
    \caption{Growth in an error of four-velocity for a trajectory generated by the final method.}
    \label{error.growth.normalization}
\end{figure}

Figure~\ref{error.growth.normalization} shows the evolution of the four-velocity error for the last method. It is clear that from the definition of the method, the four-velocity error, together with error of the integrals of motion, whose reset is part of the method, is sufficiently null after every reflection around the disk, given just by the working precision. 

\begin{figure}[H]
    \includegraphics[width=0.85\linewidth]{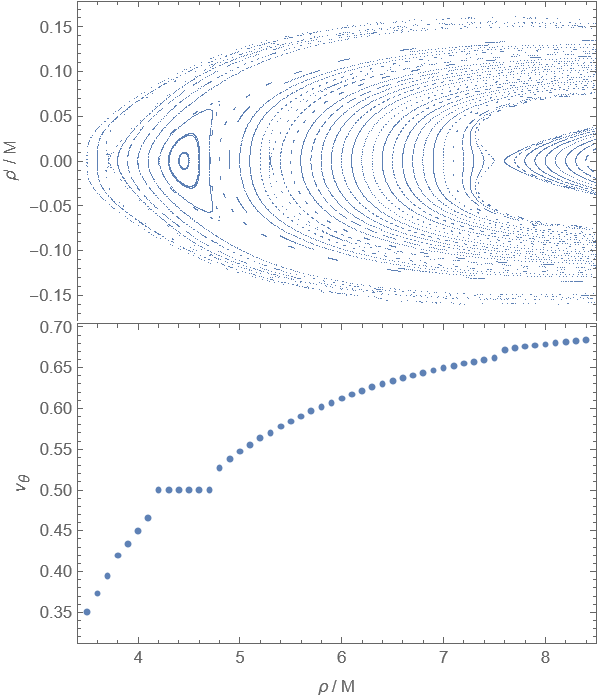}
    \caption{surface of section for the fourth method and corresponding rotation curve; parameters taken: $E = 0.95, \, L_{z} = 3.8, \, \mathcal{M}=0.2, \, b=10$  \label{rotnumb4}}
\end{figure}

Figure~\ref{rotnumb4} show a detail of the section in the range of initial conditions in $\rho$ with a corresponding rotation curve. The section is very similar to the section generated by the second method, which is understandable as the two methods share the adjustment of the position, which the first method lacks. The rotation curve is practically identical to the previous two ones, because of the averaging nature of the rotation numbers. The curve again notably depicts the plateu of constant value of $\frac{1}{2}$ attributed to the resonance $\frac{1}{2}$, disconnected from the rest of the curve and the continuous part corresponding to the KAM trajectories.

\subsection{Detail of the first island of stability}
As mentioned, Poincaré sections deform as a result of inaccuracy of the methods used. The deformation is most visible on the first island of stability, corresponding to the resonance $\frac{1}{2}$, which should be reflection symmetrical with respect to the $\dot{\rho} = 0$ axis, but its center lies visibly in the negative $\dot{\rho}$ half-plane in all Poincaré sections.

\begin{figure}[h]
    \centering
    \includegraphics[width=0.85\linewidth]{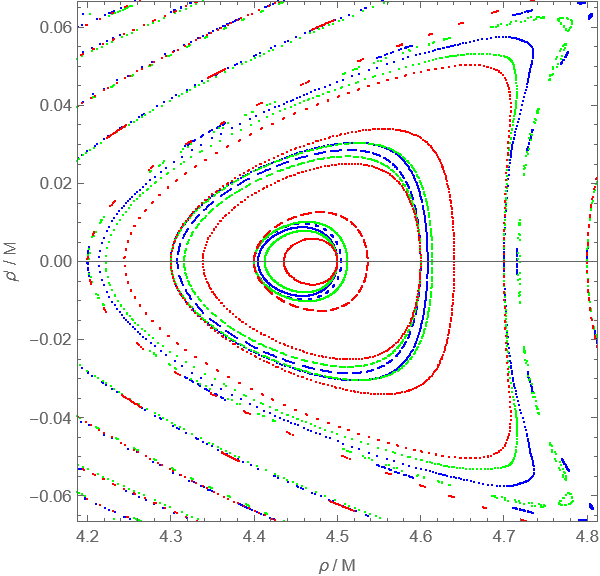}
    \caption{The first island of stability on the surface of section generated by a simple reflection (red), interpolation (green) and four-velocity renormalization (blue) method, parameters taken: $\mathcal{M} = 0.2,\, b = 10$ and $E=0.95,~L_z=3.8$} \label{fig:Pszoom}
\end{figure}

Figure~\ref{fig:Pszoom} shows the detail of the first island of stability for the first, second, and fourth method. It is visible, that the Poincaré sections do not overlap and differ by a shift in the $\rho$ position. On the basis of the trajectory continuation after the reflection and the conservation of four-velocity, it is reasonable to assume that the section produced by the fourth method is the most realistic. The second and fourth method shape more similar sections, in comparison to the first method, because of the same adjustement of position, but still differ because of the change in velocities in the fourth method.

\section{Conclusion}

In the thesis, the equations of geodesic motion were studied in the case of the system of a Schwarzschild black hole surrounded by a Vogt-Letelier disk, which is embedded in the equatorial plane. Based on the equations of motion, the phase space trajectories of a test particle were numerically simulated in Wolfram Mathematica. The singular surface, that the disk represents, had to be treated by an adaptation of the integration in proximity of the disk. A method of adaptation of the trajectory was presented and compared with the simple reflection of z coordinate in proximity of the disk, which caused the relative error to exceed $10^{-8}$, whereas the numerical precision of the integration is expected to be on the order of $10^{-16}$. The presented method consists of linear prediction of the trajectory in an area surrounding the disk and alternating corrections of four-velocity  during every reflection around the disk, where the three integrals of motion and normalization of four-velocity were used for expressing velocities $\dot{t},\,\dot{\rho},\,\dot{\phi}$. The method improved the conservation of integrals of motion to the order of $10^{-15}$. Furthermore, Poincaré sections were generated by setting them on the equatorial plane and the trajectory direction crossing from negative $z$ to positive ones. Based on the sections, rotation curves were produced. The sections respect the symmetry in $\dot{\rho}$ coordinate expected based on the space-time symmetry. The rotation curves are produced through averaging rotation angles and produce expected results, especially the plateau of value of $\frac{1}{2}$ corresponding to the first island of stability.

	\ack

O. Z. thanks the Czech Academy of Sciences for its support through the PPLZ fellowship and the Czech Ministry of Education, Youth, and Sports for its support through the LUC25003 project.
	

\def\prc{Phys. Rev. C}
\def\pre{Phys. Rev. E}
\def\prd{Phys. Rev. D}
\def\jcap{Journal of Cosmology and Astroparticle Physics}
\def\apss{Astrophysics and Space Science}
\def\mnras{Monthly Notices of the Royal Astronomical Society}
\def\apj{The Astrophysical Journal}
\def\aap{Astronomy and Astrophysics}
\def\actaa{Acta Astronomica}
\def\pasj{Publications of the Astronomical Society of Japan}
\def\apjl{Astrophysical Journal Letters}
\def\pasa{Publications Astronomical Society of Australia}
\def\nat{Nature}
\def\physrep{Physics Reports}
\def\araa{Annual Review of Astronomy and Astrophysics}
\def\apjs{The Astrophysical Journal Supplement}
\def\aapr{The Astronomy and Astrophysics Review}
\def\procspie{Proceedings of the SPIE}

\def\prc{Phys. Rev. C}
\def\pre{Phys. Rev. E}
\def\prd{Phys. Rev. D}
\def\jcap{Journal of Cosmology and Astroparticle Physics}
\def\apss{Astrophysics and Space Science}
\def\mnras{Mon. Not. R. Astron Soc.}
\def\apj{The Astrophysical Journal}
\def\aap{Astron. Astrophys.}
\def\actaa{Acta Astronomica}
\def\pasj{Publications of the Astronomical Society of Japan}
\def\apjl{Astrophysical Journal Letters}
\def\pasa{Publications Astronomical Society of Australia}
\def\nat{Nature}
\def\physrep{Phys. Rep.}
\def\araa{Annu. Rev. Astron. Astrophys.}
\def\apjs{The Astrophysical Journal Supplement}
\def\aapr{The Astronomy and Astrophysics Review}

\def\mdash{---}
	\bibliography{references}

@article{Witzany_2015,
   title={Free motion around black holes with discs or rings: between integrability and chaos – IV},
   volume={451},
   ISSN={0035-8711},
   url={http://dx.doi.org/10.1093/mnras/stv1148},
   DOI={10.1093/mnras/stv1148},
   number={2},
   journal={Monthly Notices of the Royal Astronomical Society},
   publisher={Oxford University Press (OUP)},
   author={Witzany, V. and Semerák, O. and Suková, P.},
   year={2015},
   month=jun, pages={1770–1794} }

@article{Kotlařík_2022, 
doi = {10.3847/1538-4357/ac9620},
url = {https://doi.org/10.3847/1538-4357/ac9620},
year = {2022},
month = {dec},
publisher = {The American Astronomical Society},
volume = {941},
number = {1},
pages = {25},
author = {Kotlařík, Petr and Kofroň, David},
title = {Black Hole Encircled by a Thin Disk: Fully Relativistic Solution*},
journal = {The Astrophysical Journal}
}

@textbook{Bicak:GTRtextbook,
  author    = {Jiří Bičák, Oldřich Semerák},
  title     = {Relativistic physics},
  year      = {2025},
  url       = {https://utf.mff.cuni.cz/~semerak/}
}

@article{MeissRevModPhys:1992,
  title = {Symplectic maps, variational principles, and transport},
  author = {Meiss, J. D.},
  journal = {Rev. Mod. Phys.},
  volume = {64},
  issue = {3},
  pages = {795--848},
  numpages = {0},
  year = {1992},
  month = {Jul},
  publisher = {American Physical Society},
  doi = {10.1103/RevModPhys.64.795},
  url = {https://link.aps.org/doi/10.1103/RevModPhys.64.795}
}

@article{NVoglis_1998,
doi = {10.1088/0305-4470/31/12/015},
url = {https://doi.org/10.1088/0305-4470/31/12/015},
year = {1998},
month = {mar},
publisher = {},
volume = {31},
number = {12},
pages = {2913},
author = {N Voglis and C Efthymiopoulos},
title = {Angular dynamical spectra. A new method for determining frequencies, weak chaos and cantori},
journal = {Journal of Physics A: Mathematical and General},
}

@BOOK{Contopoulos:2002ocda,
       author = {{Contopoulos}, George},
        title = "{Order and chaos in dynamical astronomy}",
         year = 2002,
       adsurl = {https://ui.adsabs.harvard.edu/abs/2002ocda.book.....C}
}

@inbook{Pound2021,
    title={Black Hole Perturbation Theory and Gravitational Self-Force},
    ISBN={9789811547027},
    url={http://dx.doi.org/10.1007/978-981-15-4702-7_38-1},
    DOI={10.1007/978-981-15-4702-7_38-1},
    booktitle={Handbook of Gravitational Wave Astronomy},
    publisher={Springer Singapore},
    author={Pound, Adam and Wardell, Barry},
    year={2021},
    pages={1–119} }

@article{Horak2003,
author = {Horák, Jiří and Krlín, Ladislav and Raidl, Aleš},
year = {2003},
month = {01},
pages = {},
title = {Deterministický chaos a jeho fyzikální aplikace /}
}
	
\end{document}